\documentclass[12pt]{article}
\usepackage{amsmath}
\usepackage{graphicx,psfrag,epsf}
\usepackage{enumerate}
\usepackage{color}
\usepackage{natbib}
\usepackage{url} 

\newcommand{\blind}{0}

\newcommand{\be}{\begin{eqnarray}}
\newcommand{\ee}{\end{eqnarray}}
\newcommand{\ba}{\begin{eqnarray*}}
\newcommand{\ea}{\end{eqnarray*}}
\newtheorem{theorem0}{Theorem}
\newtheorem{lemma0}{Lemma}
\newtheorem{remark0}{Remark}
\newtheorem{fact0}{Fact}
\newtheorem{example0}{Example}
\newtheorem{corollary0}{Corollary}
\newtheorem{proposition0}{Proposition}
\newtheorem{conjecture0}{Conjecture}
\newtheorem{definition0}{Definition}

\def\boldfacefake #1{%
    \hbox{%
        \mathsurround=0pt
        \hbox to 0.4pt{$#1$\hss}%
        \hbox to 0.4pt{$#1$\hss}%
        \hbox {$#1$}%
    }%
}
\newcommand{\nmathbf}{\bf}

\def\bfY{\mbox{$\nmathbf Y$}}

\makeatletter
\def\url@leostyle{%
  \@ifundefined{selectfont}{\def\UrlFont{\sf}}{\def\UrlFont{\small\ttfamily}}}
\makeatother

\newcommand{\go}{\rightarrow}

\newcommand{\expect}{\mbox{\rm I\kern-.20em E}}
\newcommand{\reals}{\mbox{\rm I\kern-.20em R}}
\newcommand{\sreals}{\mbox{\small \rm I\kern-.20em R}}

\begin{document}

\def\spacingset#1{\renewcommand{\baselinestretch}%
{#1}\small\normalsize} \spacingset{1}


\if0\blind
{
  \title{\bf A simple two-sample Bayesian $t$-test for hypothesis testing}
  \author{Min Wang\thanks{Corresponding author: minwang@mtu.edu}\hspace{.2cm}\\
    \small{Department of Mathematical Sciences, Michigan Technological University,
    Houghton, MI, USA}\\
    and \\
    Guangying Liu \\
   \small{ Department of Mathematics and Statistics, Nanjing Audit University,
    Nanjing, Jiangsu,  China}}


  \maketitle
} \fi

\if1\blind
{
  \bigskip
  \bigskip
  \bigskip
  \begin{center}
    {\LARGE\bf Title}
\end{center}
  \medskip
} \fi

\bigskip
\begin{abstract}
In this paper, we propose an explicit closed-form Bayes factor for the problem of two-sample hypothesis testing. The proposed approach can be regarded as a Bayesian version of the pooled-variance $t$-statistic and has various appealing properties in practical applications. It relies on data only through the $t$-statistic and can thus be calculated by using an Excel spreadsheet or a pocket calculator. It avoids several undesirable paradoxes, which may be encountered by the previous Bayesian approach of \cite{Gone:John:Lu:West:2005}. Specifically, the proposed approach can be easily taught in an introductory statistics course with an emphasis on Bayesian thinking. Simulated and real data examples are provided for illustrative purposes.
\end{abstract}

\noindent%
\vfill

\newpage
\spacingset{1.45} 

\section{Introduction}\label{section1}

In an introductory statistics course, we usually teach students how to conduct a hypothesis test based on independent samples to compare the means of two populations with equal, but unknown variance. Let $y_{ij}$ be random samples drawn from independent and normally distributed populations with means $\mu_i$ and common variance $\sigma^2$ for $j=1, \cdots, n_i$ and $i=1, 2$. We are interested in testing
\be \label{test:01}
H_0: \mu_1 = \mu_2 \quad \mathrm{versus} \quad H_1: \mu_1 \neq \mu_2.
\ee
Within a frequentist framework, the pooled-variance two-sample $t$ test is commonly used for the above hypothesis testing. The test statistic is given by
\be \label{tstat:01}
t = \frac{\bar y_1 - \bar y_2}{s_p/\sqrt{n_\delta}},
\ee
where $\bar y_i = \sum_{j=1}^{n_i} y_{ij}/n_i$ and
\be
s_p^2 = \frac{(n_1 -1)s_1^2 + (n_2 -1)s_2^2}{n_1 + n_2 -2}
\ee
is the pooled-variance estimate of $\sigma^2$ with $s_i^2 = \sum_{j=1}^{n_i}(y_{ij} - \bar y_i)^2/(n_i-1)$ for $i=1, 2$. Here, $n_\delta = (1/n_1  + 1/n_2)^{-1}$ is often called the ``effective sample size'' in the two-sample experiment. At the $\alpha$ significance level, we obtain the critical value $t_{1-\alpha/2, v}$ or P-value $p = 2P(T \geq |t|)$ with degrees of freedom $v = n_1 + n_2 -2$, where $t_{1-\alpha/2, v}$ is the $(1 - \alpha/2)$ quantile of $T_v$ distribution and $T$ has the $T_v$ distribution. We reject the null hypothesis $H_0$ if either $|t| > t_{1-\alpha/2, v}$ or $p < \alpha$; see \cite{Weis:2012}.

Bayesian approaches to hypothesis testing have recently received considerable attention and are becoming important in different disciplines, such as sociology (\citeauthor{West:1999}, \citeyear{West:1999}), economics (\citeauthor{Fern:2001}, \citeyear{Fern:2001}), and psychology (\citeauthor{Roud:Speck:Sun:2009}, \citeyear{Roud:Speck:Sun:2009}).  Many recent studies suggest that we should offer at least one course about Bayesian methods to students at early stages in their mathematics and statistics education; see, for example, \cite{Albe:1997}, \cite{Gone:John:Lu:West:2005}, \cite{Wetz:Gras:2012}, \cite{Wulf:Robin:2014}, among others. Specifically, as stated by \cite{Carl:Loui:2000}, ``\textit{The Bayesian approach to statistical design and analysis is emerging as an increasingly effective and practical alternative to the frequentist one.}'' Such a course will not only motivate students' interests in Bayesian thinking, but also help them know how to formulate Bayesian methods in simple statistical scenarios, such as the hypothesis testing in (\ref{test:01}). More importantly,  it will make students ready to use both Bayesian and frequentist ideas.

A natural approach within a Bayesian framework to compare hypotheses is the Bayes factor (ratio of the marginal densities of the two models); see \cite{Kass:95}. For the hypothesis testing in (\ref{test:01}), \cite{Gone:John:Lu:West:2005} proposed a simple closed-form Bayes factor based on the two-sample $t$-statistic and it is given by
\be \label{BF:01}
\mathrm{GBF}[H_1: H_0](\sigma^2_a) = \biggl[\frac{1+t^2/v}{1+t^2/\bigr(v(1+n_\delta\sigma_a^2)\bigr)}\biggr]^{(v+1)/2}(1+n_\delta\sigma_a^2)^{-1/2},
\ee
where $\sigma^2_a$ is a hyperparameter of the prior that needs to be specified. The choice of prior distributions for deriving the GBF will be stated in detail in the following section. The GBF in (\ref{BF:01}) shows a close relationship between frequentist and Bayesian ideas and can be easily covered in an elementary statistics course. Note that the choice of $\sigma^2_a$ is critical, because it acts as an inverse prior sample size. Specifically, the GBF with fixed $\sigma^2_a$ may exhibit some undesirable features, such as Bartlett's paradox and the information paradox; see \cite{liang:2008}. These paradoxes will definitely confuse students and even make them struggle when conducting Bayesian data analysis.

In this paper, we specify a hyper-prior for the hyperparameter $\sigma^2_a$ to reduce the impact of misspecified hyperparameter values. The prior will still result in an explicit expression of the Bayes factor based on the two-sample $t$-statistic. It is shown that the proposed approach resolves several potential difficulties and paradoxes encountered by the previous approach due to \cite{Gone:John:Lu:West:2005}. We hope that our results will facilitate an intuitive understanding and discussion of the relationship between frequentist and Bayesian ideas, but also shed some light on the importance of hyper-prior specifications to students, teachers, and researchers.

The remainder of this paper is organized as follows. In Section \ref{section:02}, we review the existing Bayes factor of \cite{Gone:John:Lu:West:2005} and discuss potential difficulties associated with fixed hyperparameter values. In Section \ref{section:03}, we specify a hyper-prior on that hyperparameter, which yields a closed-form expression for the Bayes factor. We investigate the finite sample performance of the two Bayesian procedures in a simulation study (Section \ref{section:03}) and a real-data example (Section \ref{section:04}). Some concluding remarks are given in Section \ref{section:05}, with derivation of the proposed procedure in the appendix.

\section{Bayes inference} \label{section:02}

The Bayesian analysis begins with prior specifications for the unknown parameters. Let $p(\bfY\mid \theta_j)$ and $\pi_j$ be the likelihood function of $\bfY$ and the prior probability on hypothesis $H_j$ $(\pi_0 + \pi_1 =1)$ for $j =0, 1$, respectively. From Bayes theorem, the posterior probability of $H_j$ is defined as
\begin{equation}\label{bayesfactor:1}
P(H_j\mid\bfY) = \frac{\pi_j m_j(\bfY)}{\pi_0 m_0(\bfY) + \pi_1m_1(\bfY)}.
\end{equation}
The corresponding marginal likelihood of $\bfY$ given $H_j$ is
\begin{equation} \label{likeh:01}
m_j(\bfY) = \int{p(\bfY\mid \theta_j)\pi_j(\theta_j)}\,d\theta_j,
\end{equation}
where $\pi_j(\theta_j)$ is the prior for the unknown parameter $\theta_j$ under $H_j$ for $j =0, 1$. The posterior probability of $H_1$ can be expressed as
\begin{equation}\label{bayesfactor:2}
P(H_1 \mid\bfY) = \frac{\pi_1\mathrm{BF}[H_1 : H_0]}{\pi_0 + \pi_1\mathrm{BF}[H_1 : H_0]} = \biggl[1 + \frac{\pi_0}{\pi_1}\frac{1}{\mathrm{BF}[H_1 : H_0]}\biggr]^{-1},
\end{equation}
where the Bayes factor, $\mathrm{BF}[H_1 : H_0]$, for comparing $H_1$ to $H_0$  is given by
\begin{equation}\label{m01:BF}
\mathrm{BF}[H_1 : H_0] = \frac{m_1(\bfY)}{m_0(\bfY)}.
\end{equation}
The hypothesis $H_1$ $(H_0)$ is more likely to be selected when $\mathrm{BF}[H_1 : H_0] >1$ $(< 1)$. More specifically, \cite{Jeff:1961} suggested that $\mathrm{BF}[H_1 : H_0]<0.1$, provides ``strong'' evidence in favor of $H_0$, and $\mathrm{BF}[H_1 : H_0]<0.01$, provides ``decisive'' evidence. Note that the Bayes factor for the null relative to the alternative, denoted by $\mathrm{BF}[H_0 : H_1]$, is given by
\begin{equation*}
\mathrm{BF}[H_0 : H_1] = \frac{1}{\mathrm{BF}[H_1 : H_0]}.
\end{equation*}
For the hypothesis testing problem in (\ref{test:01}), we need to specify appropriate prior distributions for $(\mu_1, \mu_2, \sigma^2)$. \cite{Gone:John:Lu:West:2005} show that this testing problem can be written in equivalent form as
\be \label{testing:01}
H_0: \delta = \mu_1 - \mu_2 = 0 \quad \mathrm{versus} \quad H_1: \delta \neq 0.
\ee
Therefore, they advocate a prior for $\delta/\sigma^2$, instead of $\mu$, where $\mu = (\mu_1 + \mu_2)/2$. After reparameterization from $(\mu_1, \mu_2, \sigma^2)$ to $(\mu, \delta, \sigma^2)$, the suggested priors are given by
\be \label{prior:01}
\pi(\mu, \sigma^2) \propto 1 / \sigma^2 \quad \mathrm{and} \quad \delta/\sigma \mid \mu, \sigma^2, \delta \neq 0 \sim N\bigl(\lambda, ~\sigma_a^2\bigr),
\ee
where $\lambda$ and  $\sigma_a^2$ are the hyperparameters that need to be pre-specified. Due to lack of prior knowledge in practice, it is natural to set $\lambda = 0$ to reflect the uncertain direction of an effect. Thus, the case for which $\lambda=0$ will be of interest to us in what follows. The Bayes factor under the above priors is
\be \label{BF:0001}
\mathrm{GBF}[H_1: H_0](\sigma^2_a) = \biggl[\frac{1+t^2/v}{1+t^2/\bigr(v(1+n_\delta\sigma_a^2)\bigr)}\biggr]^{(v+1)/2}(1+n_\delta\sigma_a^2)^{-1/2},
\ee
where  $v = n_1 + n_2 -2$. Note that the Bayes factor depends on the data only through the $t$-statistic and can often be calculated using a pocket calculator. As mentioned in the Introduction, the choice of $\sigma^2_a$ is quite critical, and in particular, the Bayes factor with fixed $\sigma^2_a$ may lead to several undesirable properties, such as Bartlett's paradox and the information paradox, briefly summarized as follows.

\noindent{\ \ Bartlett's paradox:} Because the hyperparameter $\sigma^2_a$ reflects the variance of the univariate normal distribution in (\ref{prior:01}), a large value of $\sigma^2_a$ is often chosen to minimize prior information. However, when $\sigma^2_a$ becomes sufficiently large, while $v$ is fixed ($n_\delta$ is also fixed), the GBF tends to 0, indicating that it always favors the null hypothesis, regardless of the information from the data. This phenomenon is often called Bartlett's paradox, which has been studied by \cite{Jeff:1961} and more recently by \cite{liang:2008}.

\noindent{\ \ Information paradox:} Suppose that samples are generated under $H_1$. In this setting, when $v$ is fixed, the posterior probability of $H_1$ should be higher than the one for $H_0$ when the $t$-statistic goes to infinity. We thus expect that the GBF tends to infinity as the information against $H_0$ accumulates. However, with a fixed value of $\sigma_a^2$, the GBF becomes a constant $(1+n_\delta\sigma_a^2)^{v/2}$ as $t \go \infty$. This is referred to as the information paradox.

The two paradoxes may confuse students and even make them struggle about Bayesian data analysis, especially when we introduce basic ideas of Bayesian inference in an elementary level. In this paper, we advocate a hyper-prior for $\sigma_a^2$, which not only alleviates the impacts of misspecified hyperparameter, but also yields an explicit Bayes factor. More importantly, the proposed approach is still a function of the two-sample $t$-statistic and enjoys various appealing properties, as discussed next.

\subsection{The hyper-prior for $\sigma_a^2$}

In this section, we consider a proper prior for $\sigma_a^2$, denoted by $\pi(\sigma_a^2)$. The proposed Bayes factor can be written as
\begin{equation} \label{bf:g01}
\mathrm{PBF}[H_1: H_0] = \int_0^\infty \biggl[\frac{1+t^2/v}{1+t^2/\bigr(v(1+n_\delta\sigma_a^2)\bigr)}\biggr]^{(v+1)/2}(1+n_\delta\sigma_a^2)^{-1/2} \pi(\sigma_a^2)\,d\sigma_a^2.
\end{equation}
The prior for $\sigma_a^2$ is assigned to be the Pearson type VI distribution with shape parameters $a> -1$, $b > -1$, and scale parameter $\kappa > 0$. Its probability density function (pdf) is
\begin{equation}\label{BF:fun}
\pi(\sigma_a^2) = \frac{\kappa (\kappa \sigma_a^2)^b(1 + \kappa \sigma_a^2)^{-a - b - 2}}{B(a + 1, b + 1)}I_{(0, \infty)}{(\sigma_a^2)},
\end{equation}
where $B(\cdot, \cdot)$ is a beta function. This prior has also been used by \cite{Wang:Sun:2013b} in the one-way random effects model.  With the particular choice of $\kappa = n_\delta$ and $b = (v+1)/2 - a - 5/2$, the Bayes factor can be greatly simplified as
\begin{equation}\label{BFequation}
\mathrm{PBF}[H_1 : H_0] = \frac{\Gamma\bigl(v/2\bigr)\Gamma\bigl(a + 3/2\bigr)}{\Gamma\bigl((v + 1)/2\bigl)\Gamma(a + 1)}\biggl(1 + \frac{t^2}{v}\biggr)^{(v-2a-2)/2},
\end{equation}
which is an explicit expression and can thus be easily computed using an Excel spreadsheet or a simple calculator. Such an expression is unavailable for other choices of $\kappa$ and $b$. Like the GBF in (\ref{BF:0001}), it can be regarded as a Bayesian version of the $t$-statistic; in addition, our approach enjoys several appealing properties, which are not shared by the GBF. The proof of the theorem is straightforward and is thus omitted here for simplicity.

\begin{theorem0}
In the setting of the information paradox mentioned above, the Bayes factor in (\ref{BFequation}) tends to infinity when $-1 < a < v/2 - 1$.
\end{theorem0}

The theorem shows that when $-1 < a < v/2 - 1$, the specified hyper-prior provides a resolution of the information paradox that aries in the GBF. In the case of minimum sample sizes of the two samples (i.e., $n_1 + n_2 = 3$), we have $v = 1$, indicating that $a \in (-1, -1/2)$. Of particular note is that when $a=-1/2$, the asymptotic tail behavior of
\begin{align*}
\pi\bigl(\delta/\sigma \mid \mu, \sigma^2, \delta \neq 0\bigr) = \int_0^\infty N\bigl(\delta/\sigma \mid \lambda, \sigma_a^2\bigr) \pi(\sigma_a^2) \, d\sigma_a^2
\end{align*}
becomes the Cauchy density for sufficiently large $\delta/\sigma$, which provides a flat tail behavior and diminishes the prior influence of $\pi\bigl(\delta/\sigma \mid \mu, \sigma^2, \delta \neq 0\bigr)$, especially when $a$ is small. Consequently, we recommend $a \in (-1, -1/2]$.

\begin{figure}[!htbp]
\begin{center}
\includegraphics[scale=0.5]{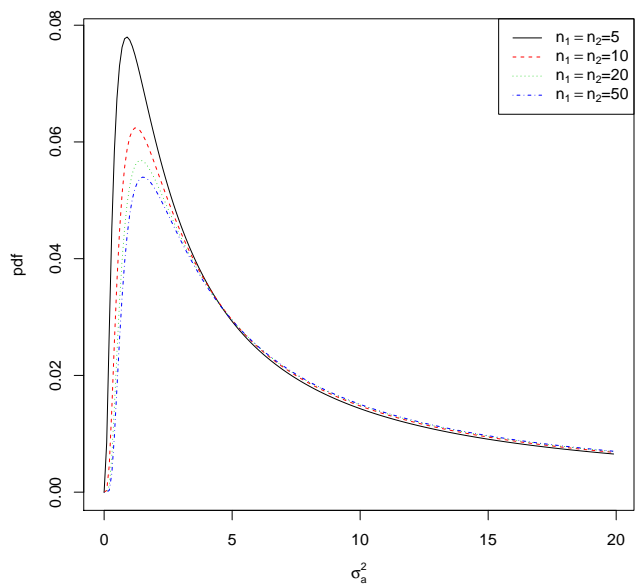}
\end{center}
\caption{The hyper-prior for $\sigma_a^2$ with $\kappa=n_\delta$, $a=-3/4$, and $b=(v+1)/2-a-5/2$ for different choices of $n_1$ and $n_2$.}
\label{priorpdf:004}
\end{figure}

It deserves mentioning that the prior depends on the sample size and that as the sample size increases, the prior has a density in the right tail that behaves like $(\sigma^2_a)^{-a-2}$, leading to a fat tail for small value of $a$. Furthermore, it can be seen from Figure \ref{priorpdf:004} that a higher prior probability is assigned to the event $\sigma_a^2 >1$. This phenomenon occurs because the parameter $\sigma_a^2$ seems to act as an inverse prior sample size. A small value of $\sigma_a^2$ (such as $\sigma_a^2\go 0$) makes the prior converge to a point mass at $\delta=0$, and the alternative $H_1$ may collapse to $H_0$. We thus obtain that the Bayes factor (the GBF) tends to 1, indicating that both hypotheses are equal descriptions to the data in the limit.

\begin{figure}[!htbp]
\begin{center}
\includegraphics[scale=0.42]{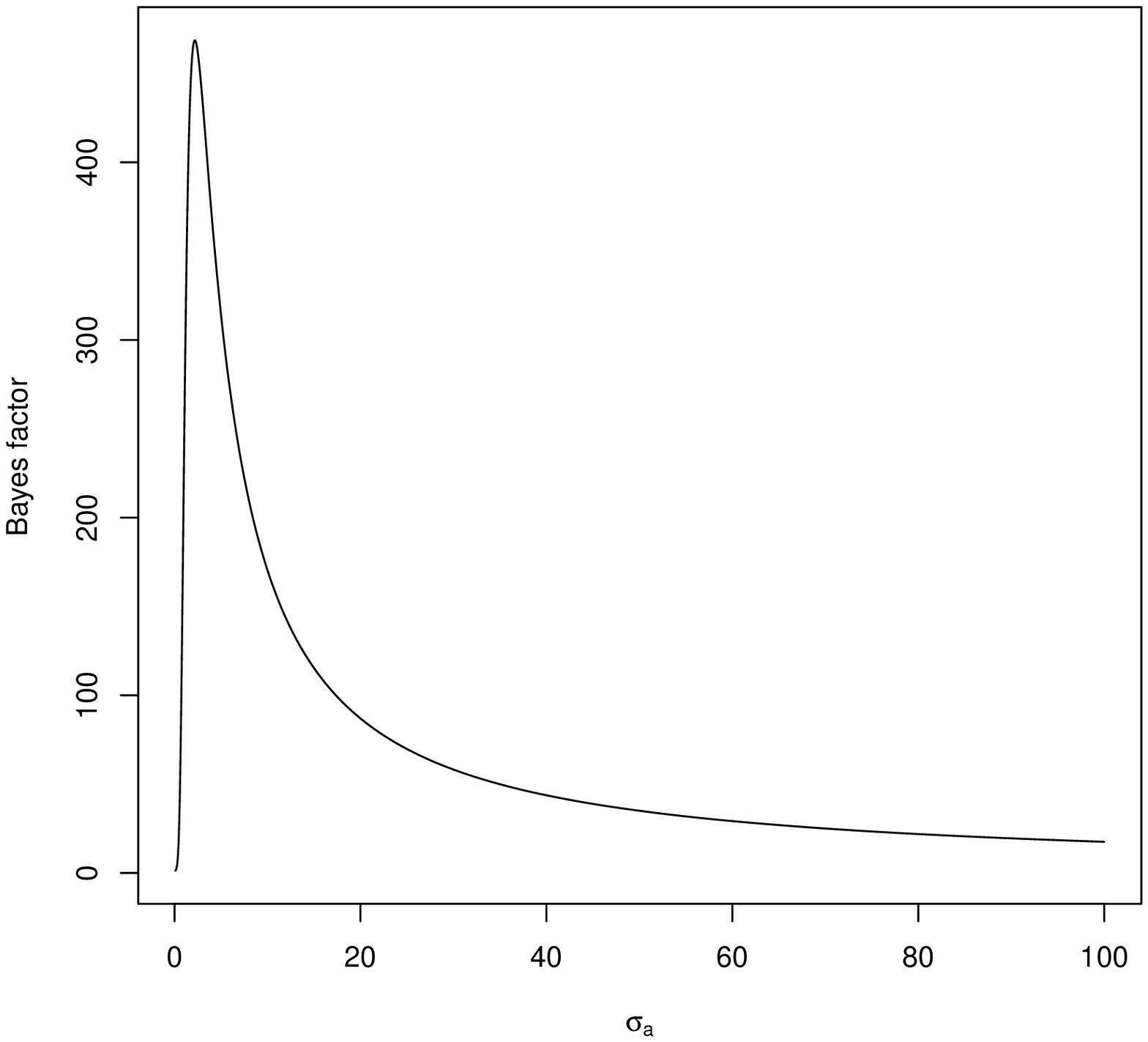}
\includegraphics[scale=0.42]{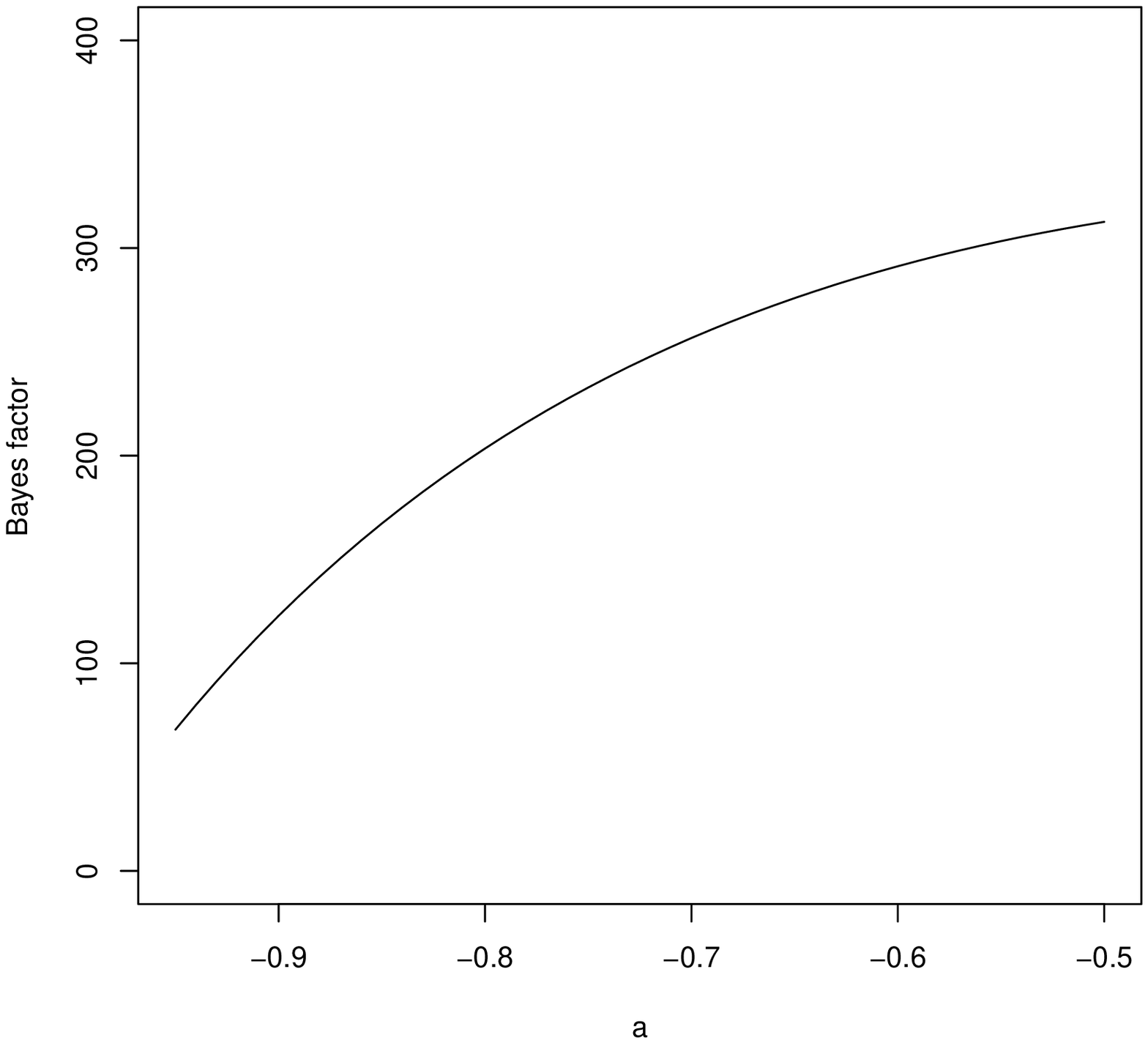}
\end{center}
\caption{The Bayes factor as a function of the hyperparameter (left: the GBF; right: the PBF) when $n_1 = n_2 =10$.}
\label{incon:01}
\end{figure}

\begin{figure}[!htbp]
\begin{center}
\includegraphics[scale=0.42]{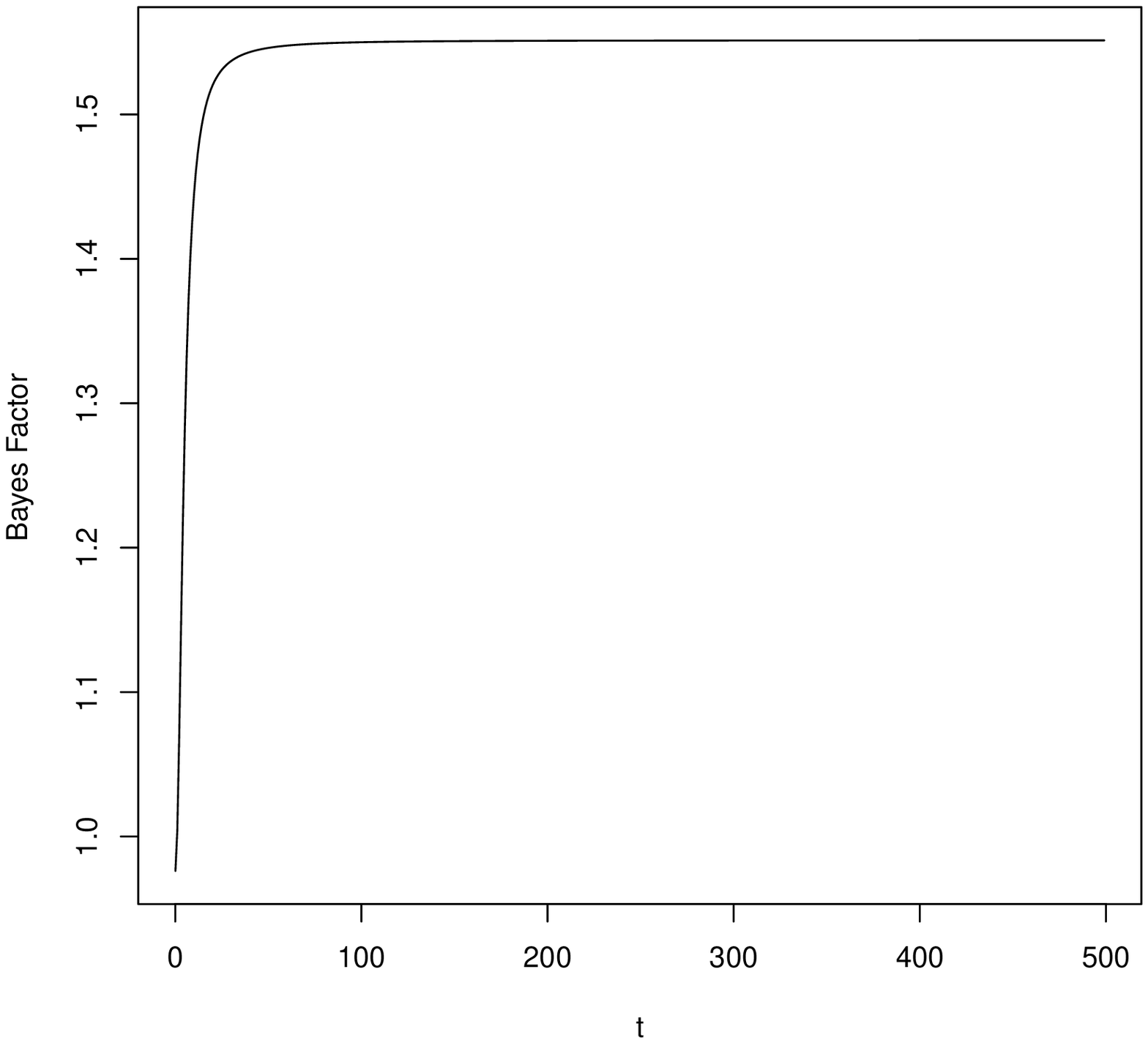}
\includegraphics[scale=0.42]{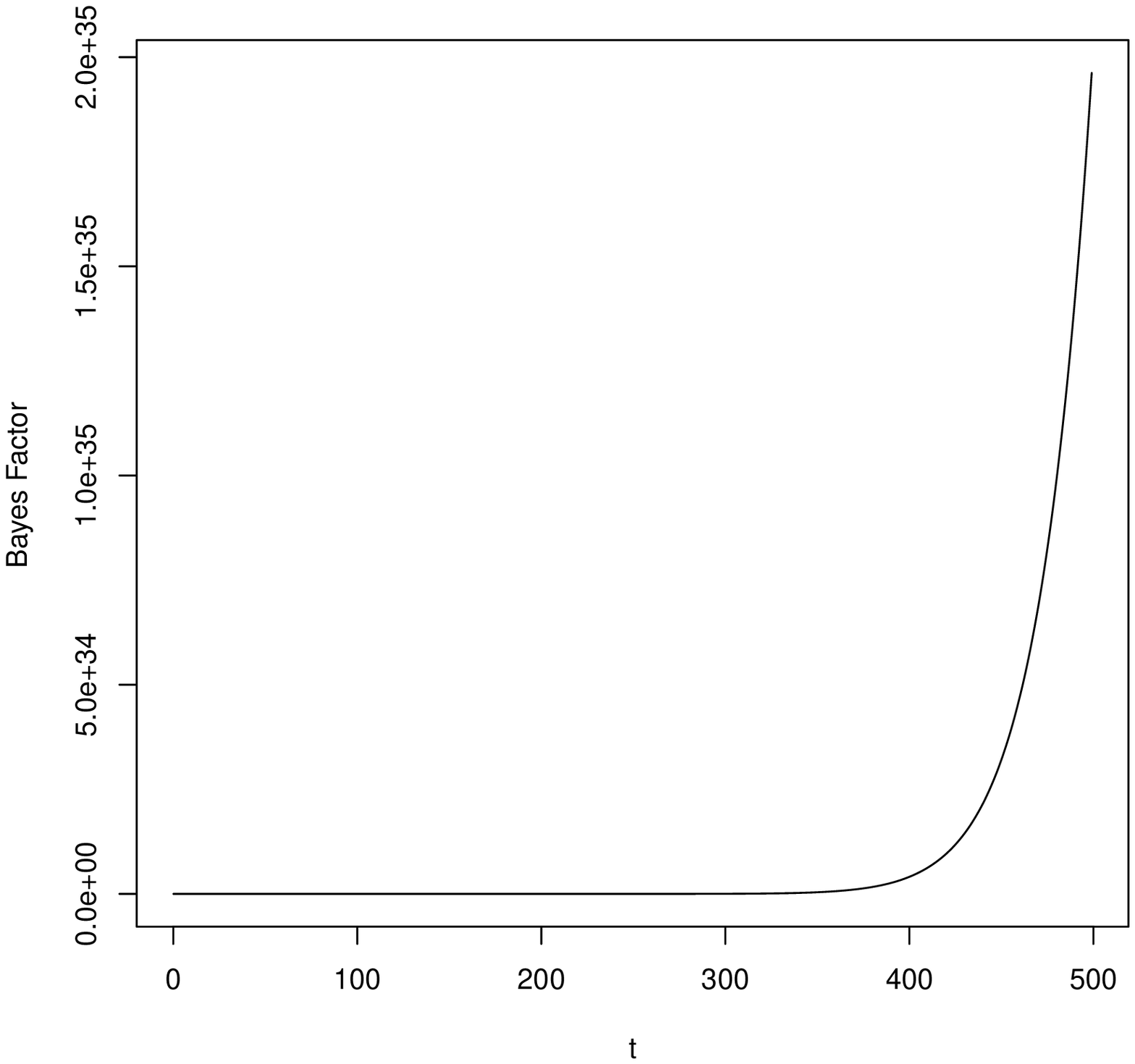}
\end{center}
\caption{The Bayes factor as a function of the $t$-statistic (left: the GBF with $\sigma_a =.1$; right: the PBF with $a=-.75$) when $t =5$ and $n_1 = n_2 = 10$.}
\label{incon:02}
\end{figure}

To see how the PBF avoids Bartlett's paradox and the information paradox, we consider two simple examples with $n_1 = n_2 =10$: one with a fixed $t$-statistic, and the other with an increasing value. Suppose that $t=5$, providing strong evidence against $H_0$. We observe from Figure \ref{incon:01} that the PBF with $a \in (-1, -1/2]$ always rejects $H_0$, while the GBF fails to reject $H_0$ when $\sigma_a$ becomes large, regardless of the information from the data. Also, it is well-known that the larger the $t$-statistic, the stronger the evidence against $H_0$. Figure \ref{incon:02} shows that as the $t$-statistic increases, the PBF grows faster than the GBF, which tends to a constant, even though $t$ becomes significantly large. These two examples show that the PBF not only avoids these paradoxes, but also provides a way to enhance students' better understanding of these paradoxes.

\section{Simulation study} \label{section:03} 

In this section, we conduct simulation studies and sensitivity analysis to investigate the finite sample performance of the two Bayes factors (GBF and PBF) with various choices of their corresponding hyperparameters. For sample 1, we generate $n_1$ random variables normally distributed with mean $0$ and standard deviation 1. For sample 2, we generate $n_2$ random variables normally distributed with mean $\delta$ and standard deviation 1, where $\delta$ ranges from $-4$ to $4$ in increments of $0.1$. To assess the sensitivity of the hyperparameters, we take $\sigma_a = \{0.1, 1/3, 0.5, 1, 1.2, 2, 5\}$ for the GBF in (\ref{BF:0001}) and $a = \{-0.95, -0.9, -0.8, -0.75, -0.7, -0.6, -0.5\}$ for the PBF in (\ref{BFequation}). For each case, we analyze $10,000$ simulated datasets with various choices of $n_1$ and $n_2$. The decision criterion used in this paper is to choose $H_1$ if the Bayes factor $>1$ and $H_0$ otherwise.

\begin{figure}[!htbp]
\begin{center}
\includegraphics[scale=0.42]{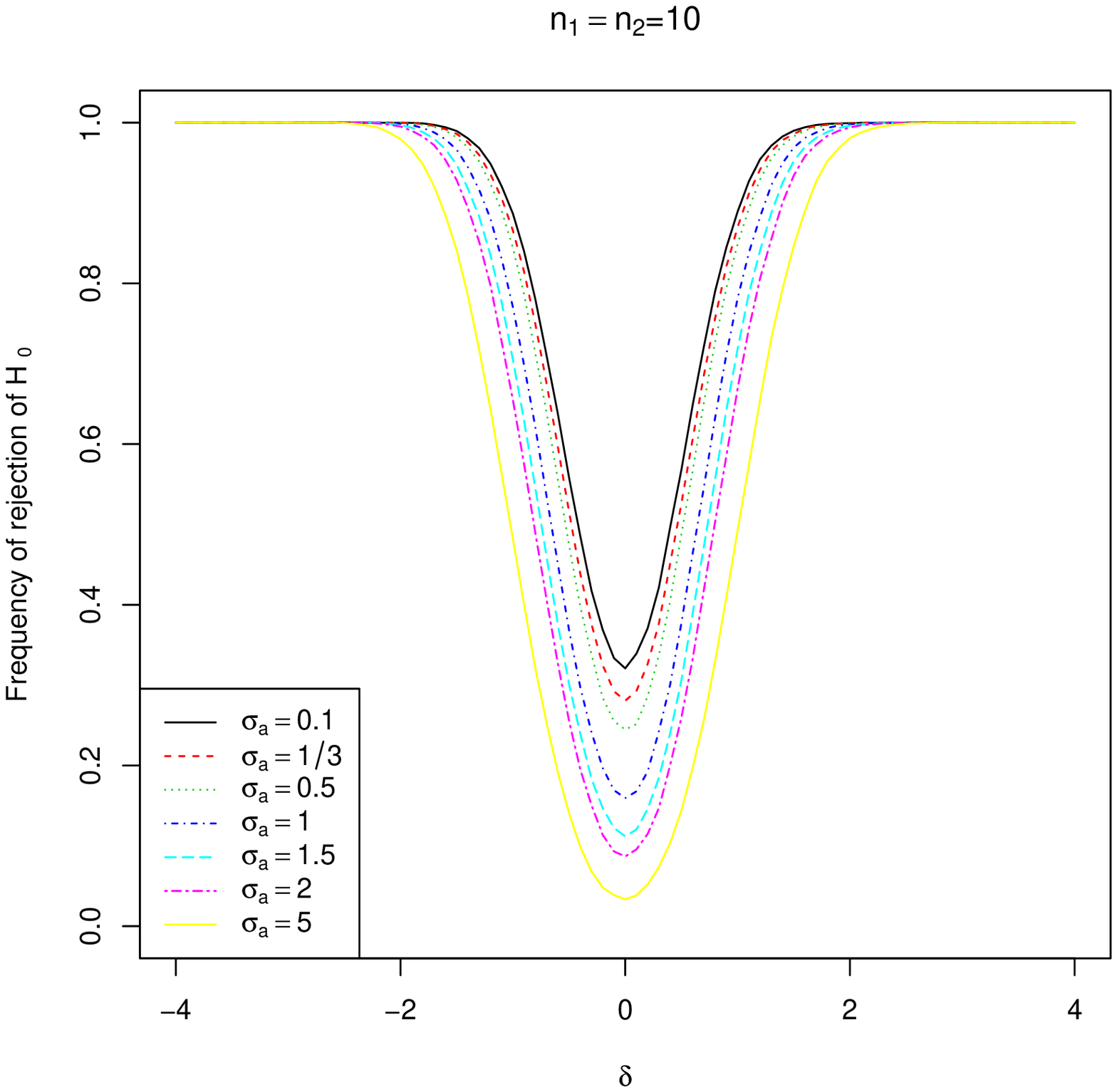}
\includegraphics[scale=0.42]{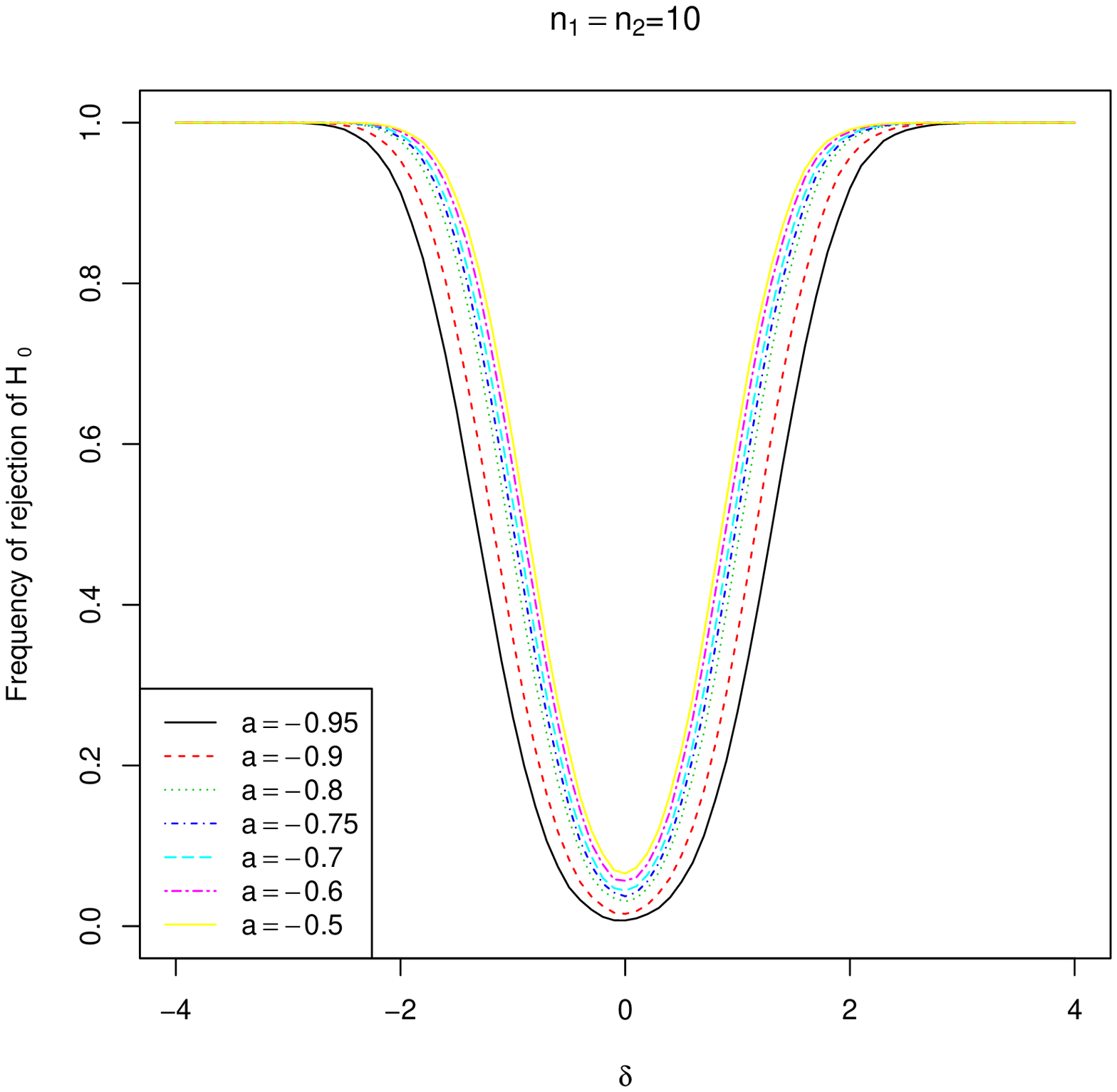}
\end{center}
\caption{The relative frequency of rejection of $H_0$ under different procedures (left: the GBF; right: the PBF) when $n_1 = n_2 = 10$.}
\label{incon:001}
\end{figure}

\begin{figure}[!htbp]
\begin{center}
\includegraphics[scale=0.42]{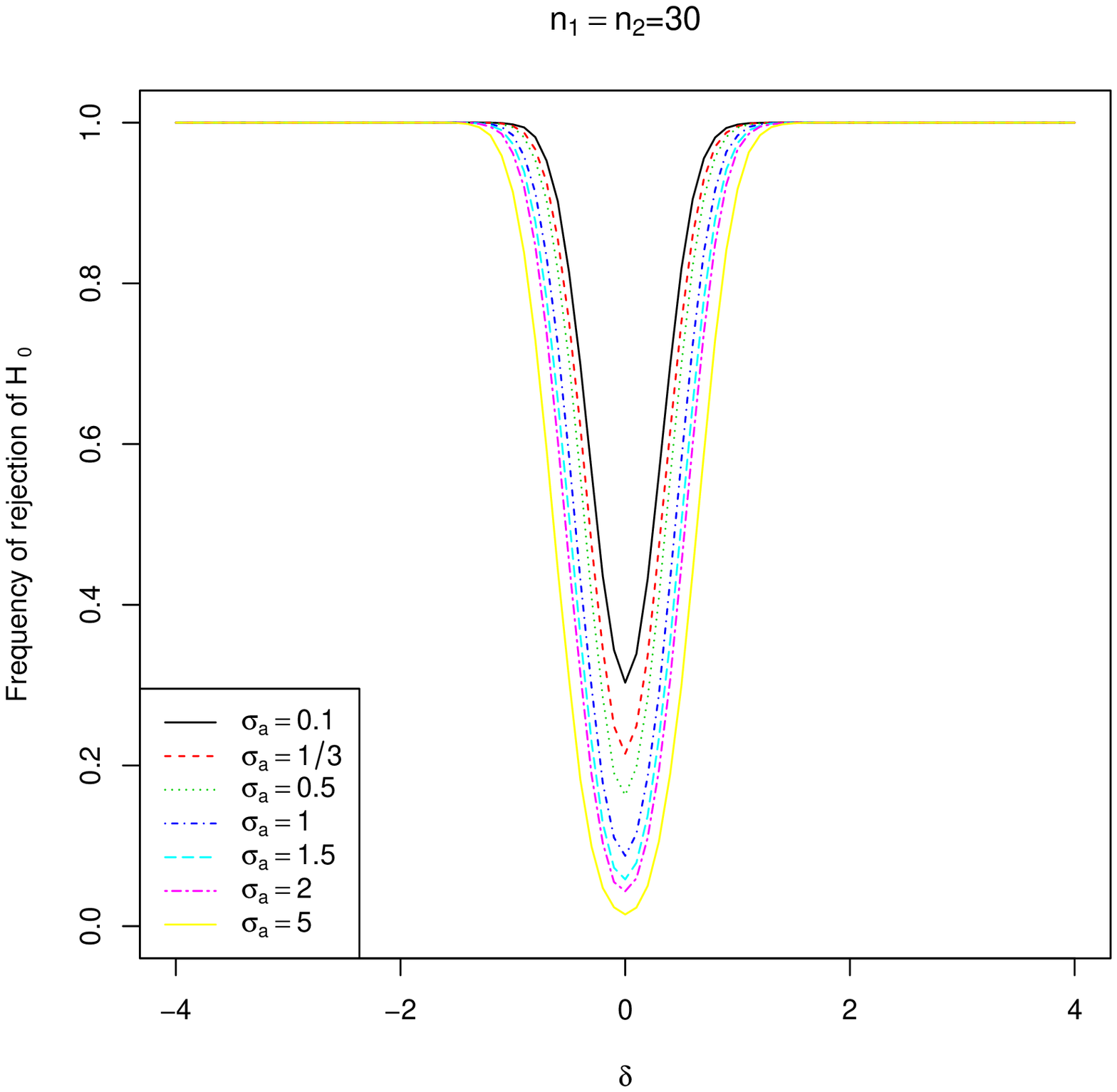}
\includegraphics[scale=0.42]{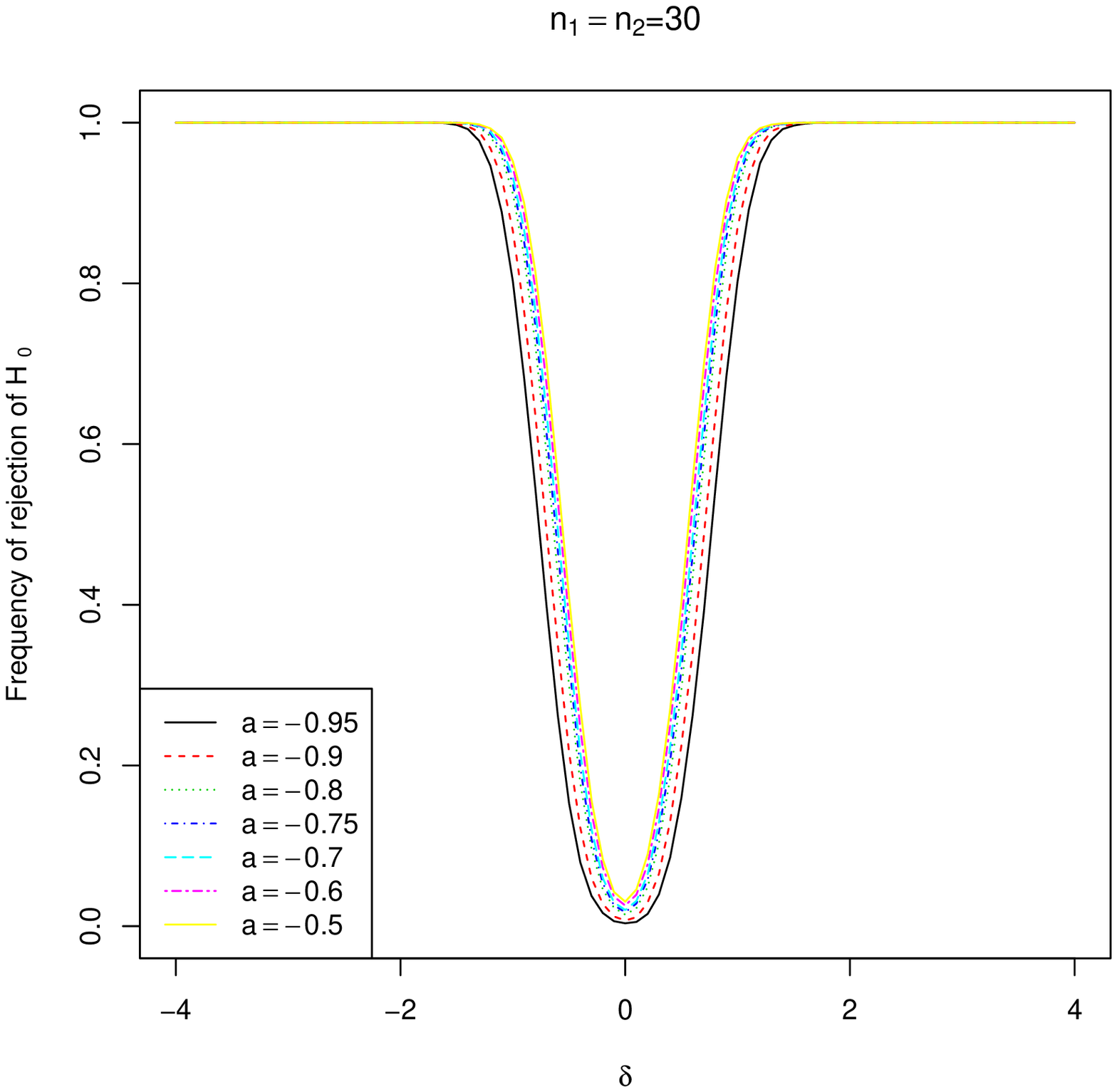}
\end{center}
\caption{The relative frequency of rejection of $H_0$ under different procedures (left: the GBF; right: the PBF) when $n_1 = n_2 = 30$.}
\label{incon:002}
\end{figure}

\begin{figure}[!htbp]
\begin{center}
\includegraphics[scale=0.42]{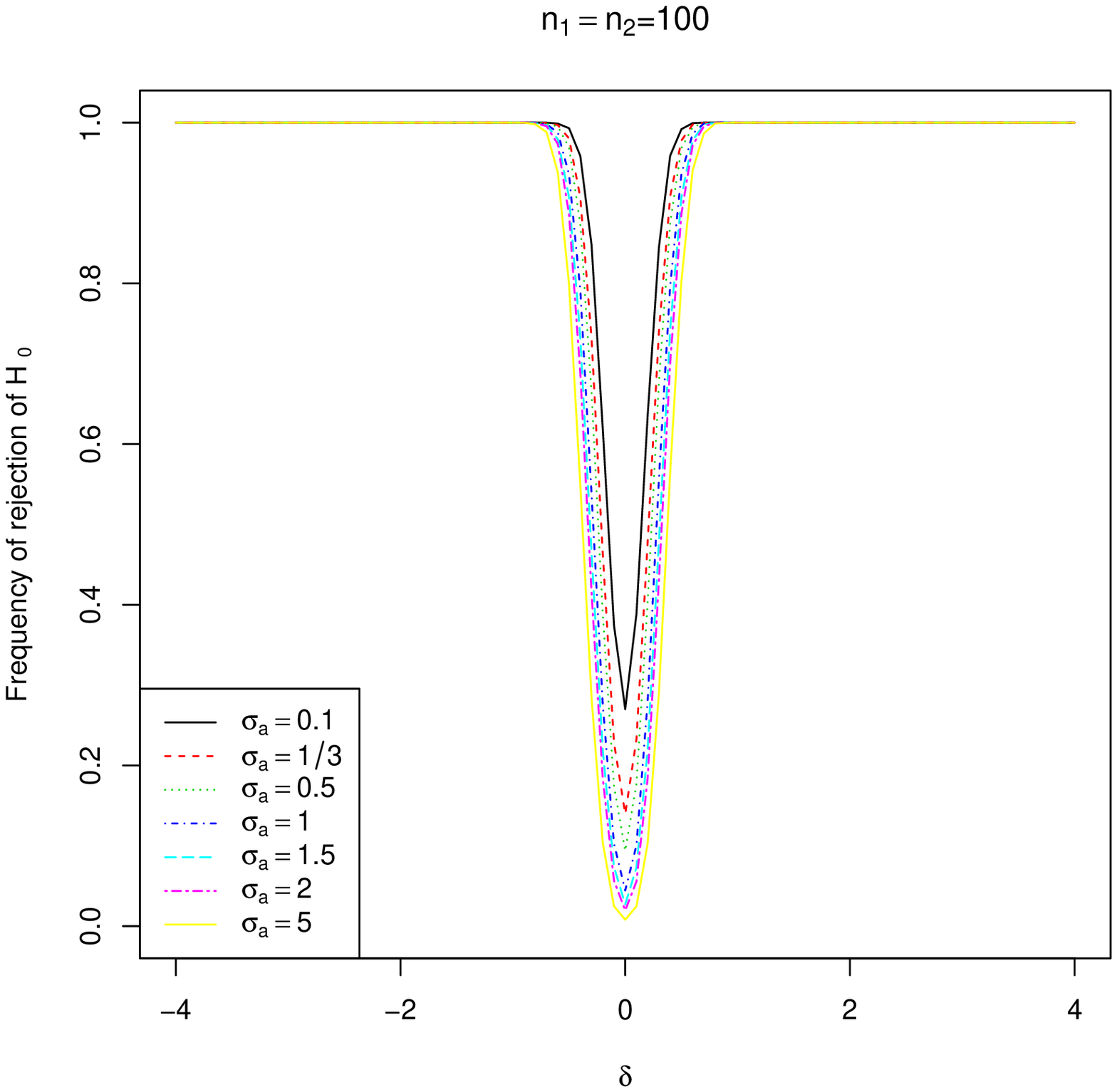}
\includegraphics[scale=0.42]{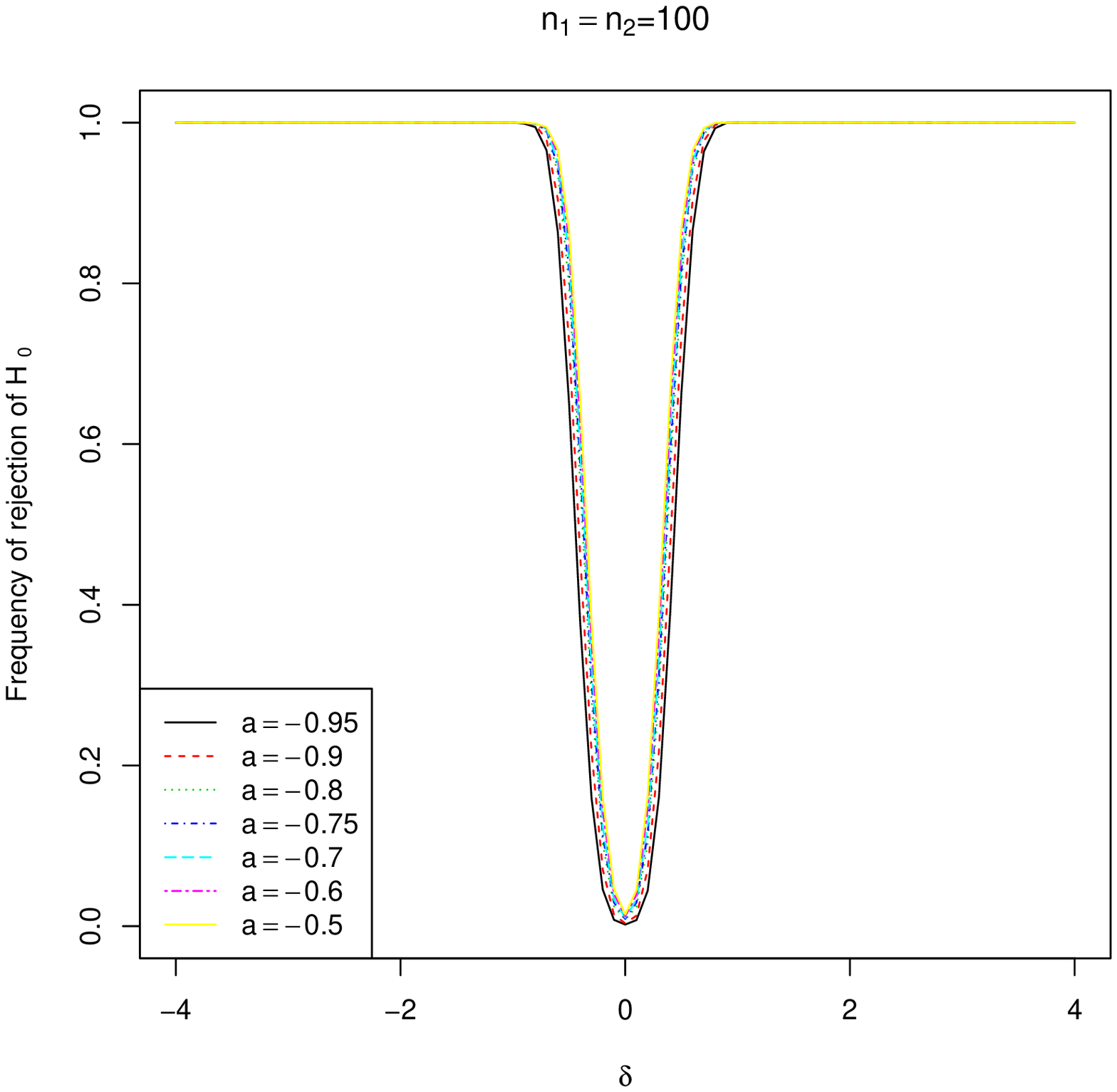}
\end{center}
\caption{The relative frequency of rejection of $H_0$ under different procedures (left: the GBF; right: the PBF) when $n_1 = n_2 = 100$.}
\label{incon:003}
\end{figure}

The relative frequencies of rejecting $H_0$ under the three different choices of sample size are depicted in Figures \ref{incon:001}, \ref{incon:002}, and \ref{incon:003}. Rather than providing exhaustive results based on these simulations, we merely highlight the most important findings from the three figures. (i) The GBF is quite sensitive to the choice of the hyperparameter $\sigma_a$, even when the sample size is large. For instance, when $n_1 = n_2 =100$ and $\delta=-0.3$, the frequency of rejecting $H_0$ changes from 0.8479 to 0.2843 with $\sigma_a$ increasing from 0.1 to 5. (ii) The PBF is relatively insensitive to the hyperparameter $a$, and when the sample size is large, the PBF behaves similarly for all values. (iii) We observe that under $H_0$ (i.e., $\delta=0$), the relative frequency of rejecting $H_0$ varies greatly for the GBF with different choice of $\sigma_a$, whereas the PBF is quite stable in terms of different value of $a$.

\begin{figure}[!htbp]
\begin{center}
\includegraphics[scale=0.42]{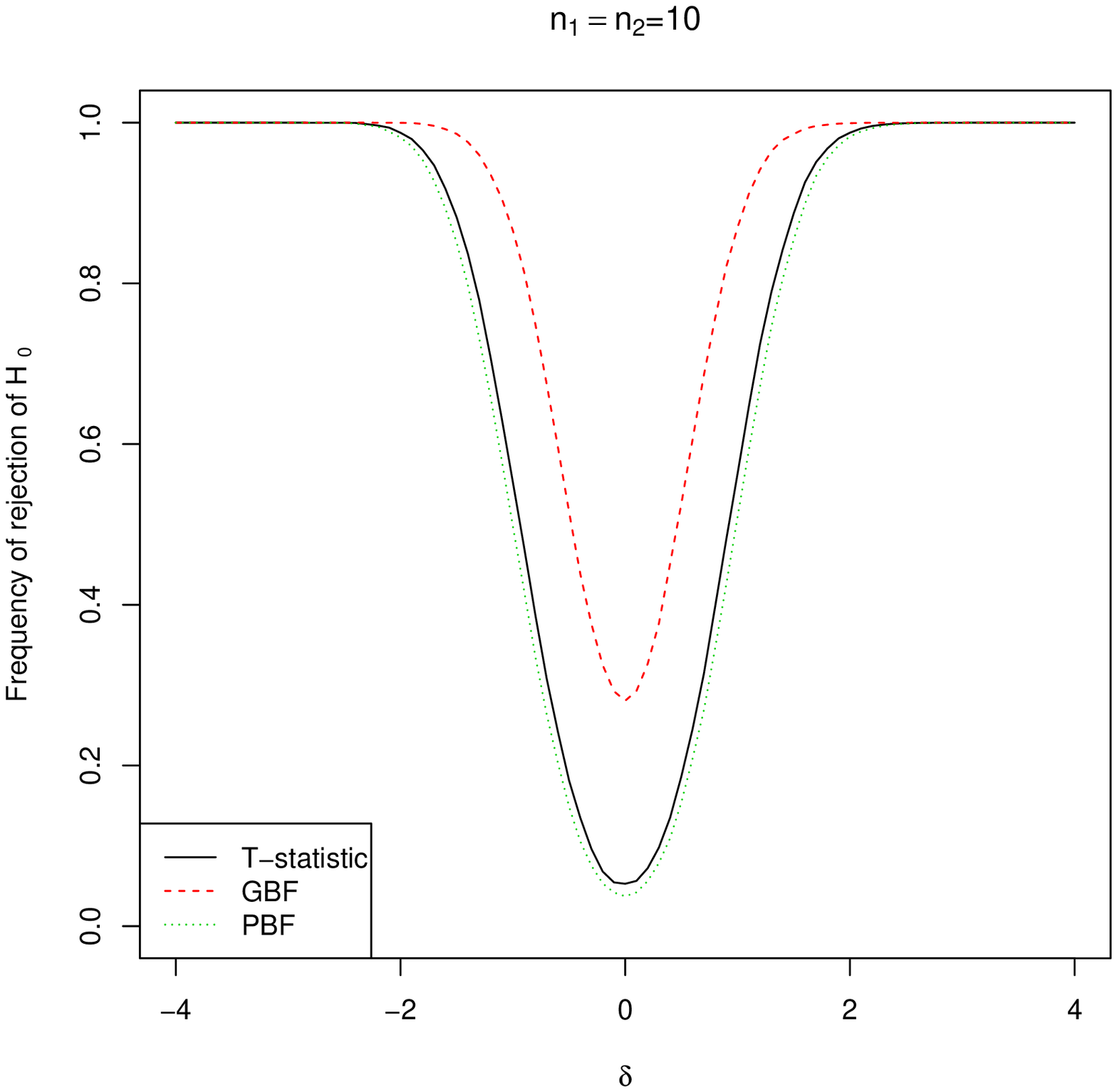}
\includegraphics[scale=0.42]{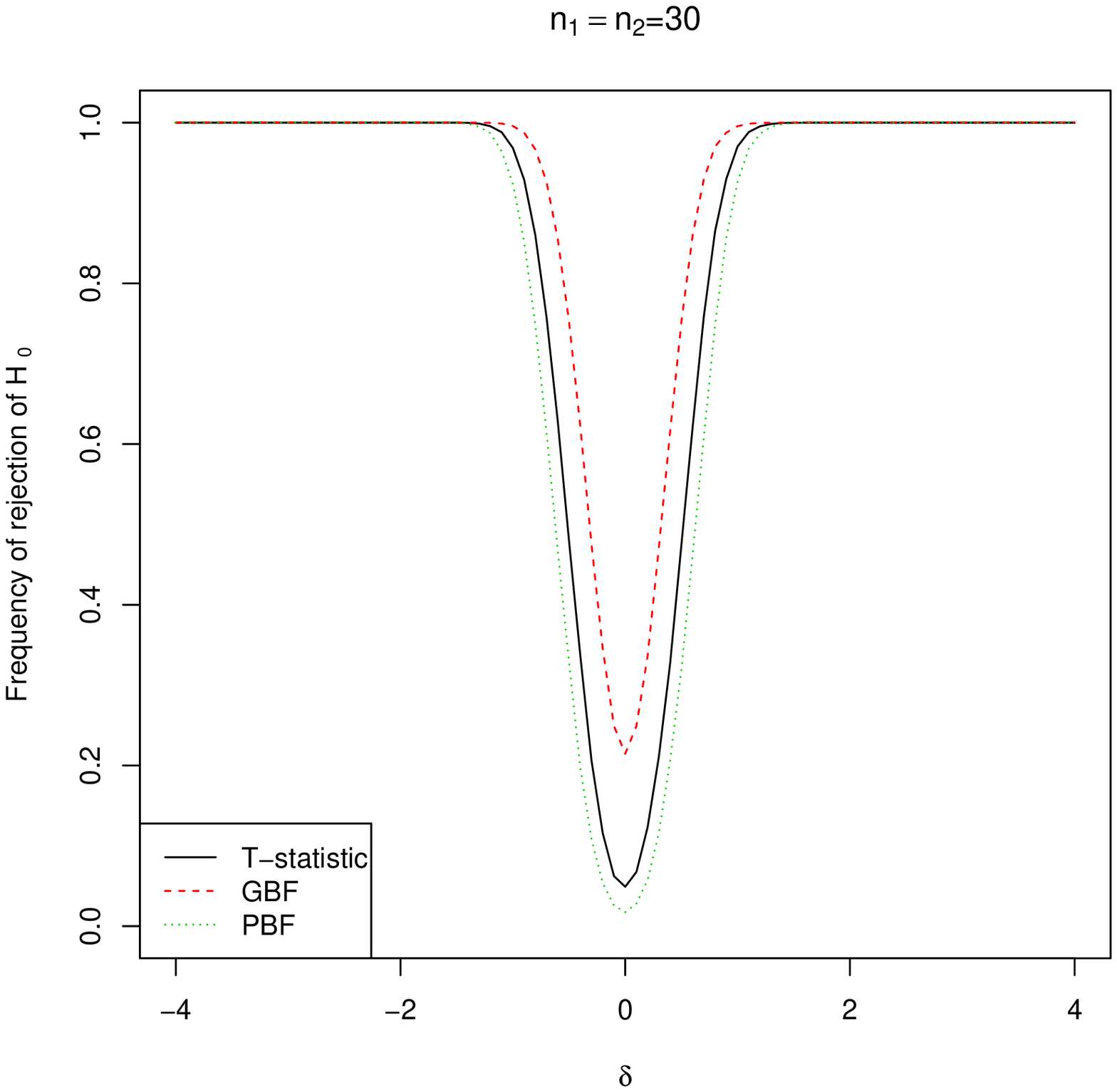}
\includegraphics[scale=0.42]{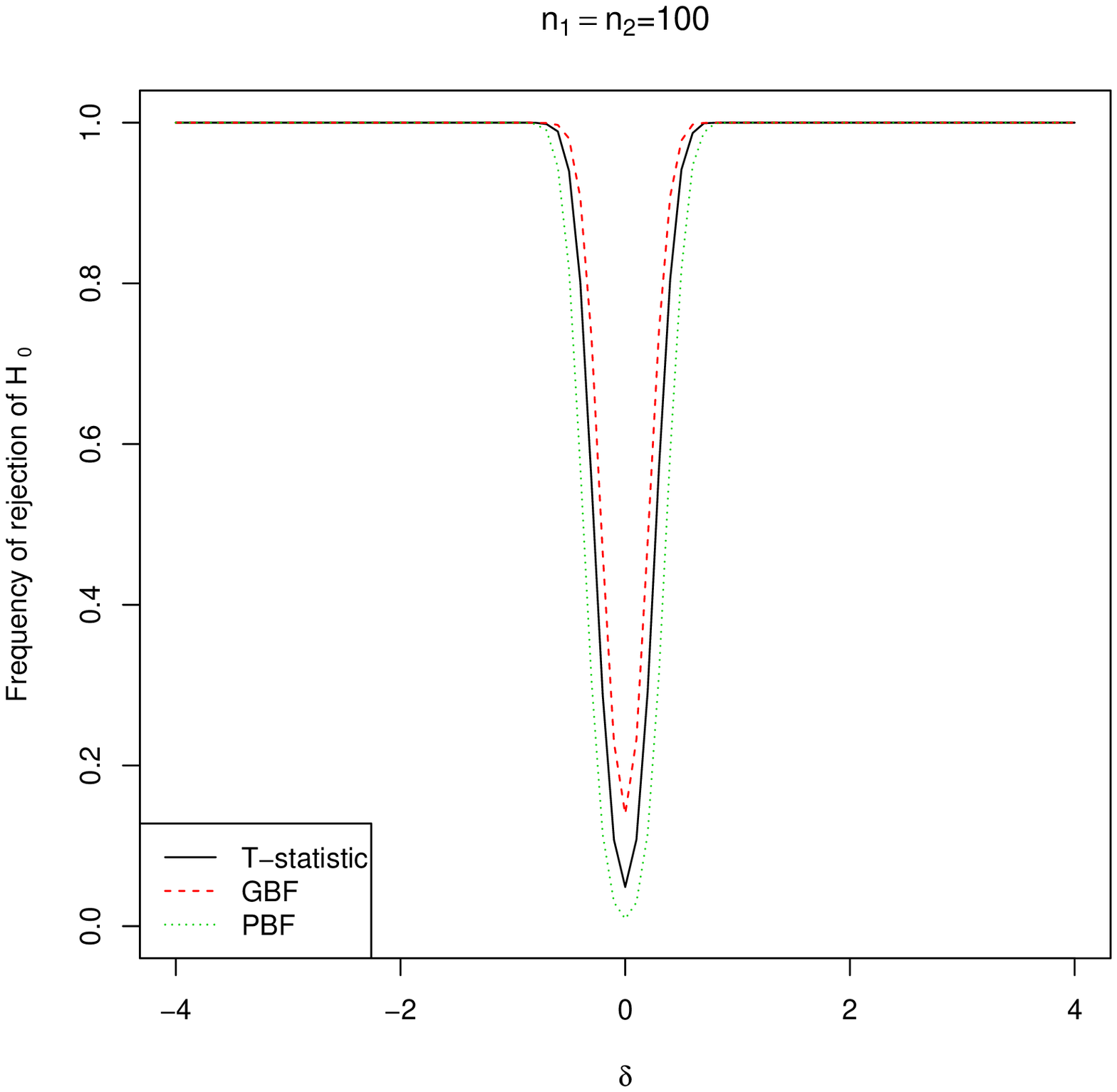}
\includegraphics[scale=0.42]{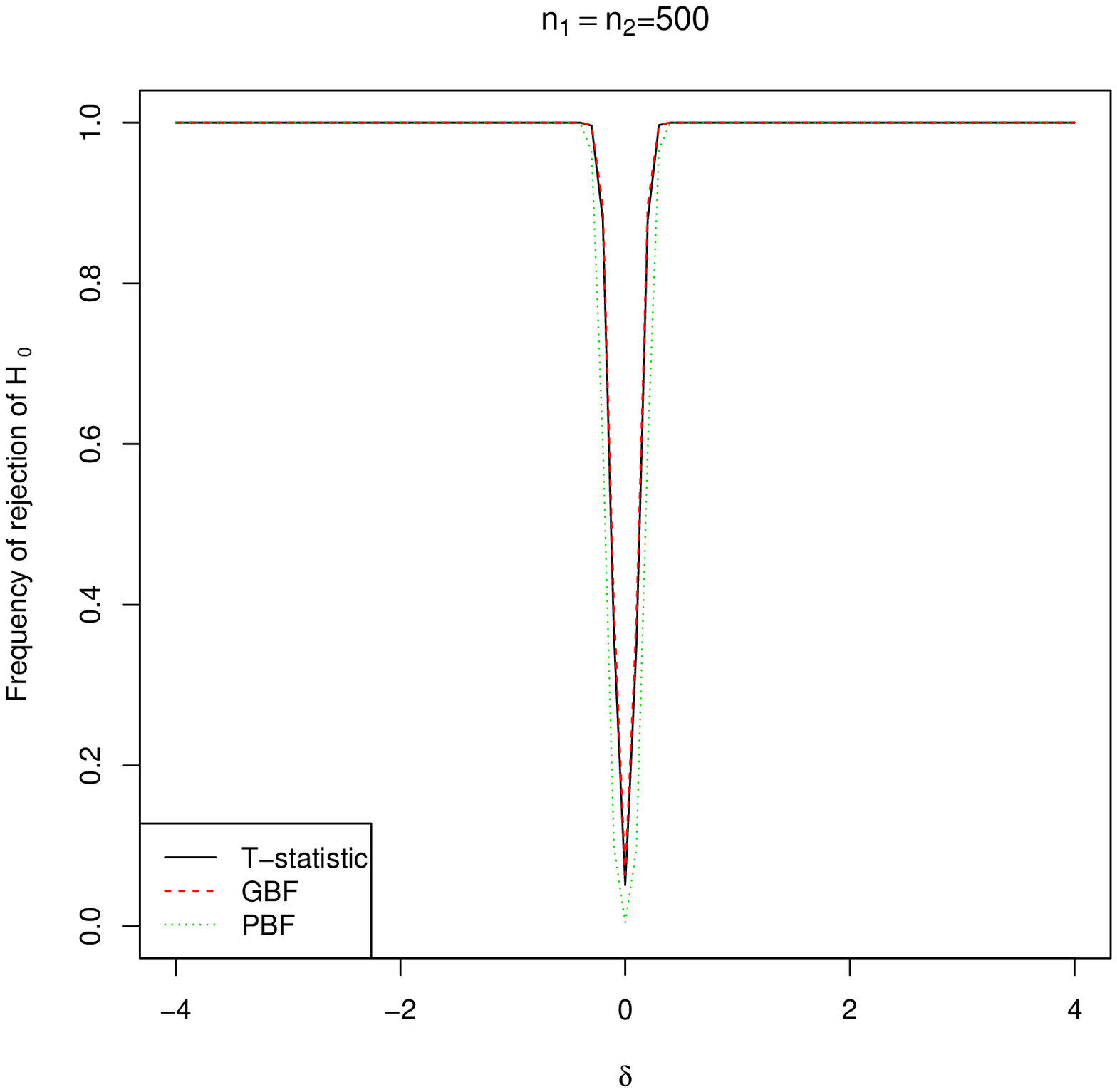}
\end{center}
\caption{The relative frequency of rejection of $H_0$ under the three considered procedures with different sample sizes.}
\label{incon:004}
\end{figure}

We now compare the performance of the two Bayes factors with the P-value based on the $t$-statistic in (\ref{tstat:01}) when $\alpha=0.05$. Based on the same simulation scheme described above, we consider the GBF with $\sigma_a = 1/3$, suggested by \cite{Gone:John:Lu:West:2005} and the PBF with $a=-3/4$. Figure \ref{incon:004} depicts the numerical findings with different sample sizes. We observe that the PBF and the P-value have similar performances, whereas they are significantly different from the GBF. As expected, when the sample size becomes large, the three procedures behave very similarly. In addition, the PBF has a faster decreasing rate to zero than the two other methods, in terms of the relative frequency of rejecting $H_0$. Thus, we may conclude that the PBF is consistent under $H_0$ when the sample size approaches infinity. This property is not shared by the two other methods under consideration.

\section{A real-data application} \label{section:04} 

We compare the performance of the two Bayes factors via a real-data example available at The Data and Story Library. ($http://lib.stat.cmu.edu/DASL/Datafiles/Calcium.html$). The data consist of the blood pressure measurements for 21 African-American men: 10 of the men took calcium supplements and 11 took placebos. We are interested in testing if increasing calcium intake reduces blood pressure.  The pooled-variance $t$-statistic is 1.634, with the two-sided P-value of 0.1187. The positive $t$-statistic indicates intake of calcium is beneficial for reducing blood pressure, and the P-value shows that the null hypothesis that calcium has no effect is more likely at the $5\%$ significance level.

To fully specify the Bayesian approach, we need to choose appropriate priors for the unknown parameters. Due to lack of prior knowledge, we consider $\pi_0 = \pi_1 = 1/2$. Therefore, for decision-making, the hypothesis $H_1$ is more likely to be selected if $P(H_1 \mid \bfY) > 1/2$, or equivalently, the value of the Bayes factor is larger than $1$.

\begin{table}[!htbp]
\centering
\begin{tabular}{c*{6}{c}c} \hline
$\sigma_a$        & 1/10   & 1/3   & 1/2   & 1     & 1.5   & 2      &  5    \\ \hline
GBF$[H_1 : H_0]$    & 1.307  & 1.264 & 1.358 & 1.193 & 0.934 & 0.746  &  0.321 \\
$P(H_1 \mid\bfY)$   & 0.509  & 0.558 & 0.576 & 0.544 & 0.483 & 0.427  &  0.243 \\ \hline
\end{tabular}
\caption{Numerical summaries of the GBF with different choice of $\sigma_a$.}
\label{table:EX1}
\end{table}

\cite{Gone:John:Lu:West:2005} analyze this dataset by using the GBF with $\sigma_a = 1/3$ and obtain that the null hypothesis is less likely because $P(H_1 \mid\bfY) = 0.558$. From a practical viewpoint, we shall be interested in a sensitivity analysis of the hyperparameter $\sigma_a$. Numerical results are reported in Table \ref{table:EX1}. We observe that as $\sigma_a$ increases, the GBF decreases. When $\sigma_a > 1$, the GBF tends to favor $H_0$, whereas  it tends to reject $H_0$ when $\sigma_a \leq 1$. The corresponding posterior probability changes from $0.509$ (against $H_0$) to $0.243$ (against $H_1$) when $\sigma_a$ changes from $1/10$ to $5$. This observation shows that the GBF is quite sensitive to the choice of $\sigma_a$ and that different choice of $\sigma_a$ may lead to a contradiction in a decision-marking process. We now employ the PBF with different values of $a \in (-1, -1/2]$. It can be seen from Table \ref{table:EX2} that the PBF is quite robust to the choice of $a$ and leads to the same decision. In addition, the conclusion based on the PBF is coincident with the one based on the two-sided P-value.

\begin{table}[!htbp]
\centering
\begin{tabular}{c*{6}{c}} \hline
$a$         & $-9/10$   & $-4/5$   & $-3/4$    & $-7/10$    & $-3/5$     &  $-1/2$    \\ \hline
PBF$[H_1 : H_0]$     & 0.177     & 0.316    & 0.375     & 0.429      & 0.534      & 0.606   \\
$P(H_1 \mid\bfY)$    & 0.150     & 0.240    & 0.273     & 0.300      & 0.344      & 0.377   \\ \hline
\end{tabular}
\caption{Numerical summaries based on the PBF with different choice of $a$.}
\label{table:EX2}
\end{table}

\section{Concluding remarks} \label{section:05}

In this paper, we propose an explicit closed-form Bayes factor for testing the difference between two means from two separate groups of subjects. The proposed approach enjoys several appealing properties. It relies on data only through the classical $t$-statistic and can thus be easily calculated using a simple calculator. It avoids several undesirable properties encountered by the approach due to \cite{Gone:John:Lu:West:2005}. More importantly, it can be easily taught in elementary statistics with an emphasis on Bayesian thinking. We hope that the results of this paper will not only facilitate an intuitive understanding of the relationship between frequentist and Bayesian ideas, but also shed some light on the importance of hyper-prior specifications to students, educators, and researchers.

\section{Appendix}

\noindent{\bf Derivation of equation (\ref{BFequation}):} When we consider the Pearson type VI distribution with $\kappa = n_\delta$, the Bayes factor in (\ref{bf:g01}) can be expressed as
\begin{align*}
\mathrm{PBF}[H_1 : H_0] &= \frac{n_\delta}{B(a + 1, b + 1)} \int_0^\infty \biggl[\frac{1+t^2/v}{1+t^2/\bigr(v(1+n_\delta\sigma_a^2)\bigr)}\biggr]^{(v+1)/2}(n_\delta \sigma_a^2)^b(1+n_\delta\sigma_a^2)^{-a-b-5/2}\,d\sigma_a^2.
\end{align*}
With the transformation $\tau = n_\delta\sigma_a^2$ and $b = (v+1)/2 - a - 5/2$, it follows
\begin{align*}
\mathrm{PBF}[H_1 : H_0] =&\frac{1}{B(a + 1, b + 1)}\int_0^\infty \biggl[\frac{1+t^2/v}{1+t^2/\bigr(v(1+\tau)\bigr)}\biggr]^{(v+1)/2}\tau^b(1+\tau)^{-a-b-5/2}\,d\tau\\[3pt]
=&\frac{(1 + t^2/v)^{(v+1)/2}}{B(a + 1, b + 1)}\int_0^\infty \biggl[1 + \frac{t^2}{v}\frac{1}{1 +\tau}\biggr]^{-(v+1)/2}\tau^b(1+\tau)^{-a-b-5/2}\,d\tau \\[3pt]
=&\frac{(1 + t^2/v)^{(v+1)/2}}{B(a + 1, b + 1)}\int_0^\infty \biggl[1 + \tau + \frac{t^2}{v}\biggr]^{-(v+1)/2}\tau^b(1+\tau)^{(v+1)/2-a-b-5/2}\,d\tau \\[3pt]
=&\frac{(1 + t^2/v)^{(v+1)/2}}{B(a + 1, b + 1)}\int_0^\infty \biggl[1 + \tau + \frac{t^2}{v}\biggr]^{-(v+1)/2}\tau^b\,d\tau ~\mathrm{since}~b = (v+1)/2 - a - 5/2\\[3pt]
=&\frac{1}{B(a + 1, b + 1)}\int_0^\infty \biggl[1 + \frac{\tau}{1 + t^2/v}\biggr]^{-(v+1)/2}\tau^b\,d\tau.
\end{align*}
With the transformation $x = \tau/(1 + t^2/v)$, it follows
\begin{align*}
\mathrm{PBF}[H_1 : H_0] = &\frac{(1 + t^2/v)^{b+1}}{B(a + 1, b + 1)}\int_0^\infty (1 + x)^{-(v+1)/2}x^b\,dx\\[3pt]
= &\frac{B(b + 1, (v+1)/2-b-1)}{B(a + 1, b + 1)}\biggl(1 + \frac{t^2}{v}\biggr)^{b+1}\\[3pt]
= &\frac{\Gamma\bigl(v/2\bigr)\Gamma\bigl(a + 3/2\bigr)}{\Gamma\bigl((v + 1)/2\bigl)\Gamma(a + 1)}\biggl(1 + \frac{t^2}{v}\biggr)^{(v-2a-2)/2},
\end{align*}
because of $b = (v+1)/2 - a - 5/2$. This completes the proof.

\bibliographystyle{annals}

\end{document}